# Delay Management Using Packet Fragmentation in Wireless Industrial Automation Systems

Anwar Ahmed Khan
*Millennium Institute of Technology & Entrepreneurship*
Karachi, Pakistan
South East Technological University
Waterford, Ireland
email: anwar.ahmed@mite.edu.pk

Shama Siddiqui
*DHA Suffa University*
Karachi, Pakistan
South East Technological University
Waterford, Ireland
email: shamasid@hotmail.com

Indrakshi Dey
*Walton Institute for Information and Communication Systems Science*
South East Technological University
Waterford, Ireland
email: indrakshi.dey@waltoninstitute.ie

*Abstract*—Managing delay is one of the core requirements of industrial automation applications due to the high risk associated for equipment and human lives. Using efficient Media Access Control (MAC) schemes guarantees the timely transmission of critical data, particularly in the industrial environments where heterogeneous data is inherently expected. This paper compares the performance of Fragmentation based MAC (FROG-MAC) against Fuzzy Priority Scheduling based MAC (FPS-MAC), both of which have been designed to optimize the performance of heterogenous wireless networks. Contiki has been used as a simulation platform and a single hop star topology has been assumed to resemble the industrial environment. It has been shown that FROG-MAC has the potential to outperform FPS-MAC in terms of energy efficiency and delay both, due to its inherent feature of interrupting ongoing lower priority transmission on the channel.

## I. Introduction

In the context of modern industrial automation and control systems, advanced technologies, such as wireless communication, artificial intelligence, big data analytics, digital twins and blockchains have played a pivotal role. The advent of Industry 4.0 and Industrial Internet of Things (IoT) has ushered in countless opportunities for enhancing industrial output by minimizing malfunctions and reducing downtime [1]. Amidst these advancements, managing heterogeneous traffic emerges as a core challenge for wireless industrial automation systems, demanding efficient handling of prioritized traffic streams. As industries strive for seamless operations and optimized performance, addressing this challenge becomes imperative to ensure the smooth functioning of critical processes and systems [2]. Hence, exploring innovative techniques for delay management becomes increasingly significant for industrial wireless networks.

For the wireless communication, delay management can be implemented at various layers of communication stack. Since MAC layer plays a vital role in scheduling of packets, it has been one of the hottest areas of interest of past researchers to develop priority mechanisms [3]. Some of the major techniques proposed for ensuring quick delivery of high priority data include adaptive contention windows [3], queue management [4], adaptive data rate adjustment [5], duty-cycle adaptation [6], wake-up radio [6] and multi-channel usage [8]. Also, various hybrid schemes have been proposed, which combine some of the advanced techniques along with using conventional super-frame method [9] and Time Division Multiple Access (TDMA). Although most of these techniques guarantee a prioritized channel access for high priority traffic, if the lower priority traffic has already started to transmit, it is no longer possible for the higher priority traffic to interrupt the ongoing transmission. Moreover, for the super-frame based protocols which claim slot-stealing, stealing is only possible until the transmission is scheduled, not yet begun [9].

In industrial settings, wireless sensors may generate heterogeneous traffic of varying priorities in several applications. For example, wireless sensors deployed to monitor the health and performance of machinery may generate heterogeneous traffic. Critical equipment, such as turbines or pumps, might require real-time monitoring and immediate response to prevent breakdowns, while less critical equipment may have lower priority traffic for periodic status updates. Similarly, industrial facilities often utilize wireless sensors to monitor environmental conditions, such as temperature, humidity, and air quality. Certain parameters, like temperature in a furnace or chemical concentration in a storage tank, may require immediate attention in case of deviations, while others, such as ambient humidity, may have lower priority. Moreover, wireless sensors used for safety and security purposes, such as fire detection systems, intrusion alarms, or access control systems also generate heterogeneous traffic [11]. Critical events like fire or unauthorized access may require immediate notification and action, while routine security checks may have lower priority. Clearly, most of the mentioned scenarios represent an emergency and it is not practical for the urgent data generated during this to wait for the completion of ongoing transmission on the network, if any.

In this work, we advocate for the usage of packet fragmentation scheme, FROG-MAC [10] for the industrial automation system. FROG-MAC is a MAC protocol which has specifically been designed for prioritized heterogenous traffic and can be used for a wide range of applications, such as wireless body area networks, vehicular ad hoc networks and industrial networks. In this protocol, the traffic of varying priority attains a varying level of channel access. The higher priority traffic gets an immediate access through interrupting the ongoing transmission of lower priority packets. This is ensured by transmitting the lower priority packets in the form of short fragments, while the higher priority traffic is always transmitted as a single unit. Further details of FROG-MAC's operation will fall later in the paper. We compare the performance of FROG-MAC with FPS-MAC (which is a fuzzy priority scheduling based MAC) in this work, specifically for industrial sensor systems.

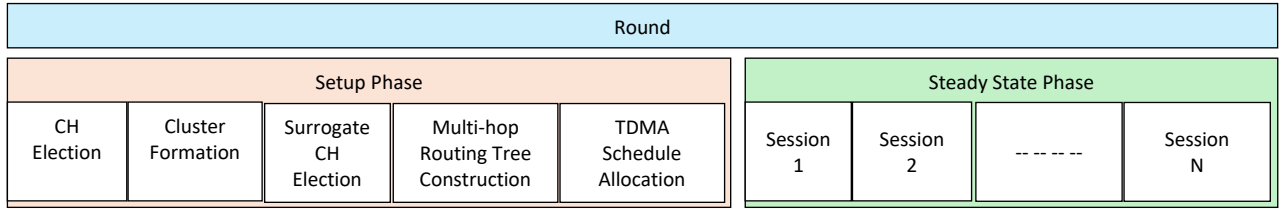

Figure 1: Operation Phases of FPS-MAC [11]

Rest of this paper has been organized as follows: Section II summarizes the relevant work; section III presents the experimental settings; section IV details the results and discussion and finally, section VI concludes the work and suggests future work directions.

## II. RELEVANT WORK

In this section, we present a brief overview of FPS-MAC and FROG-MAC, the two protocol which have been compared for their performance in terms of latency for the industrial sensor networks. FPS-MAC works on the principles of slot stealing and fuzzy based scheduling, whereas the core concept of FROG-MAC is data fragmentation.

FPS-MAC has been designed to steal the data slots from periodic traffic in order to transmit the higher priority data first [11]. This way, fuzzy priority scheduling is done in the event-based scenarios to guarantee the timely access for emergency traffic and also, to ensure the appropriate level of Quality of Service (QoS). FPS-MAC operates in two phases of set-up and steady state phase, as shown in Figure 1. In the setup phase, the Cluster Head (CH) selection and cluster formation take place. To ensure that each cluster remains operational even after the failure of Master Cluster Head (MCH), the surrogate CH election is also conducted. Moreover, the routing tree is created and TDMA schedule is allocated. Subsequently, both the intra- and inter-cluster data transmission takes place during the steady state phase. As shown in Figure 2, TDMA frames are differently designed to manage the periodic monitoring and emergency situations.

For the event situation, the Emergency Indication Slot (EIS slot) is used, where the nodes having some urgent data indicate the channel requirement. These nodes must listen for the EIS period, and if they find the channel idle, they can begin transmitting the indication signal. Here, there are two possible transmitter nodes for the indication signal; a node which has just detected an event, or a node which has to transmit previously buffered event traffic. The CH then acknowledges the indication signal and switches the transmission mode from periodic to emergency; this is followed by the control period during which all the member nodes remain active in order to obtain information about the current TDMA frame. Later, there are some operational differences for the frames as shown in Figures 2 (a) and (b), based on whether the protocol has to deal with routine or emergency traffic. Fuzzy based scheduling is another major contribution of FPS-MAC, where the priority level of each node is computed using information about "intra-cluster distance (distance between member node and Cluster Head (CH))," "residual energy," "slots required," and "emergency bit".

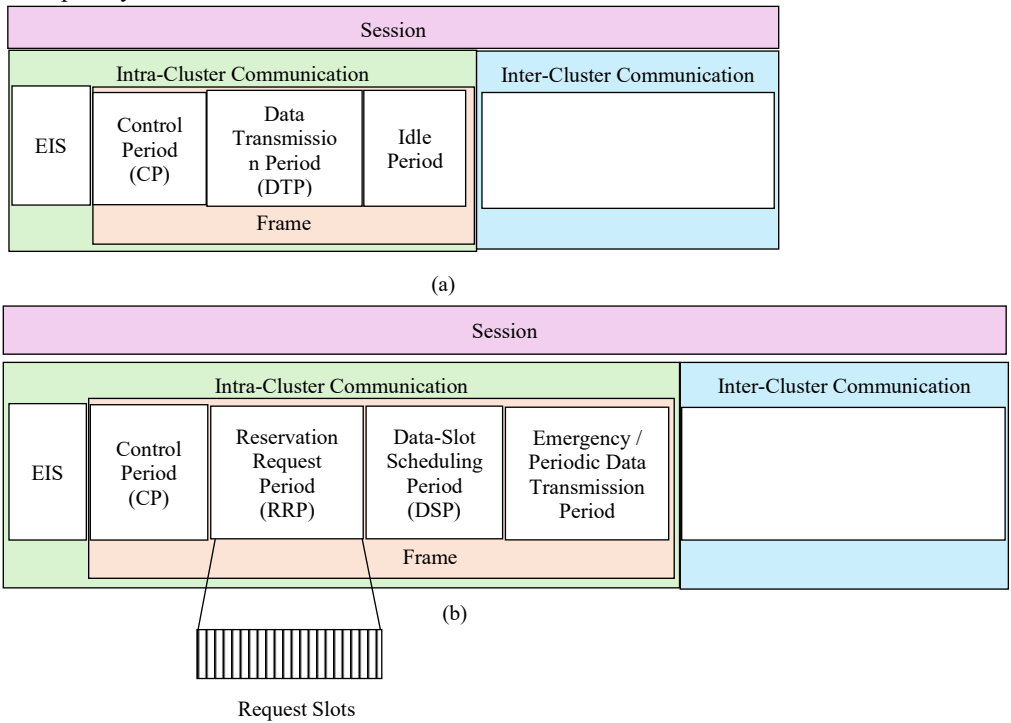

Figure 2: TDMA Frame Format for FPS-MAC- (A) Periodic Transmission, (B) Event Situation [11]

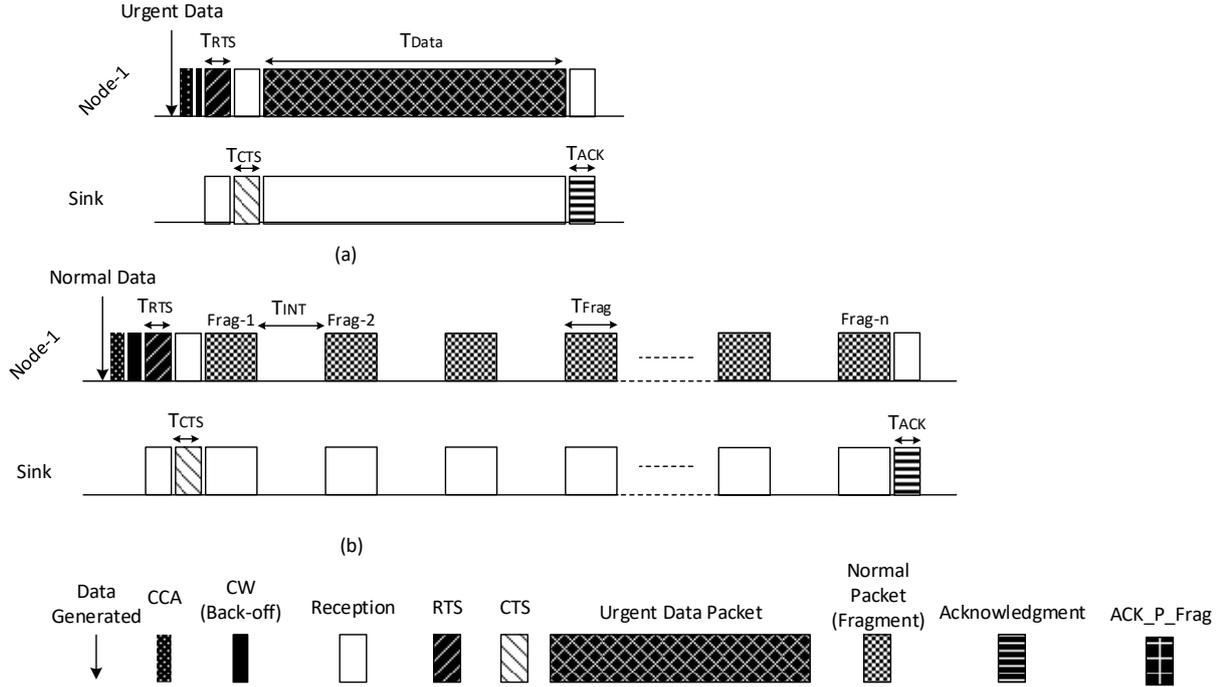

Figure 3: Basic Operation of FROG-MAC [13]

FROG-MAC [10] is an asynchronous MAC protocol where the operation has been defined on the basis of heterogenous priority data. FROG-MAC introduced a groundbreaking technique: the ability to interrupt ongoing transmissions on the channel—a feat previously unattainable with existing MAC schemes, despite the development of numerous priority-based MAC protocols tailored for mission-critical applications [12]. The basic operation of FROG-MAC is shown in Figure 3, where the low priority data transmission is done in fragments (Figure 3a), and high priority data is sent as a single packet (Figure 3b). The pauses between fragment transmissions allow the data of emergency nature to request and obtain channel access quicker, as compared to if had to wait for the complete transmission of ongoing lower priority data.

### III. EXPERIMENTAL SETUP

The experiments are performed for testing the service provided to two levels of services generated by the nodes in a single hop network, emergency and normal. The emergency traffic represents data of time-critical and unpredictable nature such as fire occurrence or gas leakage. On the other hand, normal data represents the packets periodically generated to communicate the health and state of plants/equipment; this category of traffic will include examples of temperature and humidity data.

20 nodes were set in a star topology, whereas 21st node acted as the cluster head/sink. The number of nodes sending emergency traffic varied between 3 and 18 assuming the spread of event, and its detection by various nodes; similar assumption has been made in [11], where it is stated that in the case of fire occurrence, various nodes will continue to detect and report as the fire spreads. Star topology is used in this paper, assuming that all the nodes will be reporting emergency event to their cluster head. However, for more complex topologies, such as linear multi-hop, experiments can be performed using multi-priority data, where fragmentation would be valuable for reducing the delay.

TABLE I.   SIMULATION SETTINGS

| Simulation Parameter | Simulation Settings |
|---|---|
| Simulation Area | 50 X 50 m |
| Simulation Duration | 5000 Sec |
| Total Number of Nodes | 21 |
| Number of Transmitting nodes | Variable |
| Message Generation Interval of Urgent/Emergency/Event-detection Traffic | 2 min |
| Message Generation Interval of Normal/Periodic Traffic | 10 sec |
| Data Packet Length | 34 Bytes |
| Fragment Size for FROG-MAC | Varying (2 to 32) |

For simulating the FPS-MAC, the nodes used dynamic TDMA based scheme as discussed earlier, whereas fragmentation was implemented on the normal data for simulating FROG-MAC. Other simulation settings used for the present study are shown in Table 1.

### IV. RESULTS AND DISCUSSION

The delay performance comparison of FPS-MAC and FROG-MAC has been illustrated in Figure 4, by varying the fragment size. For FPS-MAC, the delay has been higher as compared to FROG-MAC because of the underlying differences in the TDMA-based and asynchronous protocols. In FPS-MAC, the lower priority nodes have to wait for their allocated slot for transmission, which could be few sessions away; same is the case for event-detecting nodes, which might not always get the channel access during their first attempt during EIS period. On the other hand, in FROG-MAC, the nodes get a chance to transmit without waiting for the typical handshaking being performed in the FPS-MAC (the schedule information communication between the CH and each node). For the FROG-MAC, there is a slight difference in delay when number of fragments are increased; this is because of the possibility of interruption by higher priority traffic during the transmission, and also because of the additional header bytes that will be sent along with the excessive fragment transmissions.

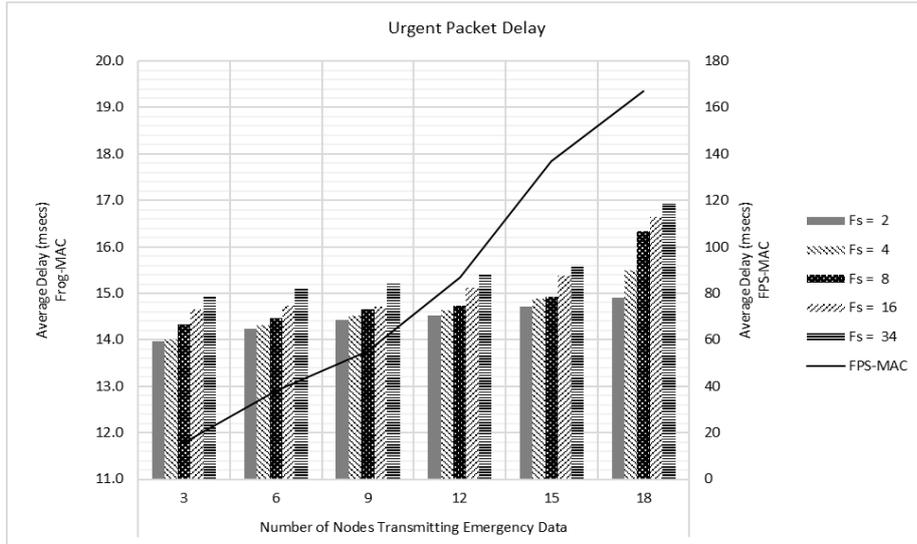

Figure 4: Average Delay Comparison of FROG-MAC and FPS-MAC by Varying Number of Transmitting Nodes and Fragment Size

Next, the delay performance is compared for the emergency/urgent traffic by varying the number of nodes that send urgent data; the results have been shown in Figure 5; here, the fragment size for FROG-MAC was chosen as 8 bytes. For the increasing number of nodes, the delay is shown to be rising for both protocols, as there will be a higher contention between event-detecting nodes. However, since the waiting time is much lower for FROG-MAC, we see a significant difference in the delay results. Firstly, the nodes operating with FROG-MAC do not have to wait for the EIS; secondly, there is even a possibility of getting channel busy during EIS for FPS-MAC; thirdly, once a node sends an indication message during EIS, it has to wait for its turn; there could also be a probability that it does not receive an acknowledgement from the CH, which would imply that the node might have to wait for another EIS period despite having won the channel in the first attempt. On the other hand, the nodes in FROG-MAC only have to wait for the short fragment being transmitted on the channel; as soon as it is done, the nodes which detect event could quickly grasp the channel and send their data. Finally, the priority assignment in FPS-MAC is done based on the fuzzy algorithm, which might not always result in true representation of emergency identification. On the other hand, for now, the FROG-MAC has been assigned well-defined static priorities which would ensure that higher priority nodes always get access to channel by interrupting the lower priority data.

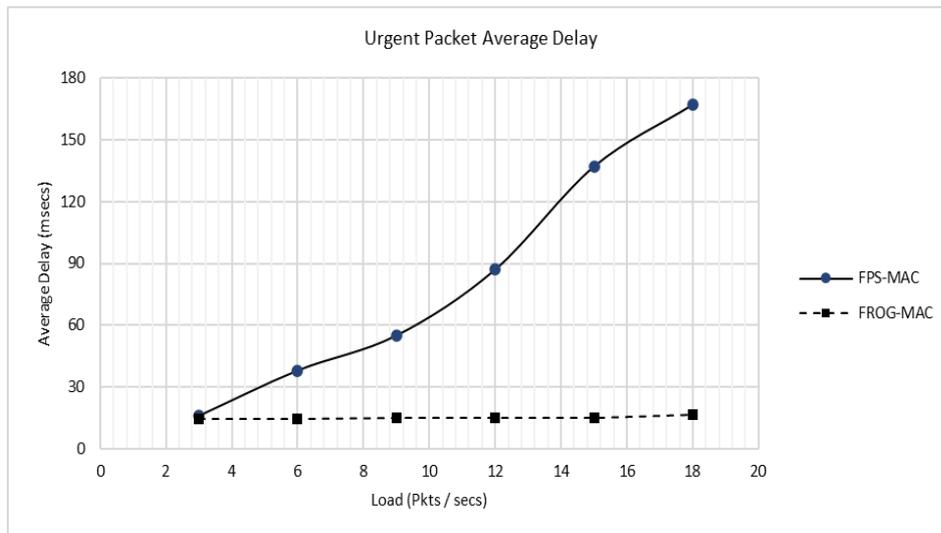

Figure 5: Average Delay Comparison of FROG-MAC and FPS-MAC by Varying Traffic Load

## V. Conclusion and Future Work

We presented a performance comparison of MAC protocols designed for dealing with prioritized heterogenous traffic, which is common in industrial environment today. We chose FPS-MAC, where the nodes steal slots for facilitating the higher priority traffic, and FROG-MAC where the lower priority data is fragmented in order to provide early channel access to the urgent traffic. We varied the number of nodes transmitting urgent data and overall traffic load to represent the industrial data, and compare the 2 protocols. Moreover, the impact of fragmentation has also been illustrated. It has been found that FROG-MAC outperforms FPS-MAC in terms of latency, due to providing the chance of interruption of ongoing transmission, which is not possible in the operation of FPS-MAC.

In future, we plan to enhance the functionality of FROG-MAC by integrating it with the machine learning algorithms. The protocol will be designed to learn from the previous operational cycles so the optimal fragment size could be decided for each type of traffic. This would ensure achieving an even higher level of performance. Moreover, we also plan to focus on enhancing the reliability and robustness of FROG-MAC to ensure uninterrupted communication in challenging industrial environments. This includes mitigating packet loss, minimizing interference, and implementing error detection and correction mechanisms to maintain data integrity and reliability under adverse conditions. Also, FROG-MAC will be compared with standard protocols for various applications, such as with IEEE 802.11-p for vehicular networks.

## Acknowledgment

This contribution is supported by HORIZON-MSCA-2022-SE-01-01 project COALESCE under Grant Number 10113073.